\documentclass[]{aipproc}
\layoutstyle{8d}

\usepackage{color}
\usepackage{axodraw}
\usepackage{fancybox}
\usepackage{psfrag}
\usepackage{epsfig}
\usepackage{amsfonts}
\def\bea{\begin{eqnarray}}
\def\eea{\end{eqnarray}}
\def\be{\begin{equation}}
\def\ee{\end{equation}}

\def\Z{{\bf Z}}

\begin{document}

\title{The Higgs boson as a gauge field in extra dimensions}{}

\author{Marco Serone}{address={SISSA, Via Beirut 2, I-34014 Trieste, Italy},}


\begin{abstract}

I review, at a general non-technical level,
the main properties of models in extra dimensions where
the Higgs field is identified with some internal component
of a gauge field.
\end{abstract}

\maketitle

The Standard Model of fundamental interactions (SM) represents the
best theory at our disposal to describe all high-energy processes we know so far.
Most likely, however, it cannot be the final description of Nature.
Gravity is excluded from the theory, the origin of many parameters, such as the Yukawa couplings,
is unexplained and it is also afflicted by hierarchy problems.
The latter are best understood if one considers the SM as an effective field theory
valid up to energy scales of order $\Lambda$, above which the theory has to be replaced by a more
fundamental (and yet unknown) microscopic theory.
At the quantum level, one finds that two parameters in the SM heavily depends on the details
of the microscopic theory: the cosmological constant and the Higgs mass.
For instance, by using a simple cut-off regularization at the
scale $\Lambda$, one finds that the radiative corrections to the
cosmological constant and the Higgs mass are respectively proportional to the quartic and quadratic power
of $\Lambda$. We do not know the value of $\Lambda$, but the phenomenological success of the SM puts a
bound on it: $\Lambda_{Exp} \geq$ few Tev (see {\em e.g.}
ref.\cite{unknown:2003ih}).\footnote{A remark is in order here. More precisely,
the bound is on certain higher derivative operators, suppressed by powers of $\Lambda$. The value reported
assumes that the coefficients of the operators are of order one. If this is not the case,
$\Lambda$ does not necessarily coincide with the scale at which new physics arises.}
Leaving aside the outstanding problem of the cosmological constant hierarchy problem (whose solution might
well be due to unknown quantum gravity effects), we have still to face
the problem of why and how the electroweak scale (and thus the Higgs mass) is stabilized
to a value which is roughly one order of magnitude smaller than the minimal experimentally
allowed value for $\Lambda$. Sometimes one refers to this problem as the
gauge ``little hierarchy problem''. Although it involves only one order of magnitude,
one has to notice that, contrary to the ``usual'' gauge hierarchy problem in which one takes
$\Lambda \sim M_{Planck}$, this is an experimental fact and it does not rely on any
assumption about the scale at which new physics should appear.

During the years, many solutions have been proposed to address the gauge hierarchy problem.
Independently of the precise nature of the Higgs field that is assumed in each of these
proposals, all of them require, in one way or another, the
appearance of new physics at $\Lambda\sim$ TeV.

The Minimal Supersymmetric Standard Model (MSSM) is at the moment
the best candidate theory of new physics beyond the Standard Model. However, no super particle
has been discovered yet and, as far as the little hierarchy problem is concerned,
the MSSM needs some unwanted fine tuning. It is thus important to investigate alternative
scenarios where radiative corrections to the Higgs mass can be somehow suppressed.

Theories formulated in $D>4$ space-time dimensions seem to be a promising
arena for new ideas along this direction.
Being non-renormalizable, these theories must always be seen as effective theories
valid up to an UV cut-off scale $\Lambda$ (not to be confused with the SM cut-off introduced before),
above which the extra dimensional theory
needs an UV completion.
It is in particular important to have an estimate of $\Lambda$ in order to quantify
the relevance of quantum corrections given by higher derivative operators and
understand the energy range of validity of the effective theory.
A good estimate of $\Lambda$ (which is typically hard to determine otherwise)
is obtained by using Na\"ive Dimensional Analysis (NDA).
The low-energy effective theory is trustable
only if $\Lambda \gg 1/L,E$, where $L$ is the typical size of the compact extra dimensions
and $E$ is the energy of the process under consideration.

There are several ideas and theoretical frameworks in the context of extra dimensions.
We focus here on the idea that the SM Higgs boson arises from
the internal component of a higher-dimensional gauge field of a group
$G\supset G_{SM}=SU(3)_c\times SU(2)_L\times U(1)_Y$.
By choosing suitable gauge groups in the extra dimensions, one can incorporate
all SM gauge bosons ($\gamma$, $W^{\pm}$, $Z$ and gluons) and the Higgs field $H$
as arising from different components of the same higher dimensional gauge field
$A_M$, where $M$ runs over all (usual and extra) space-time coordinates.

Due to this common origin of the gauge and the Higgs fields, this idea is sometimes
called ``gauge-Higgs unification''.
Its essential point is that the Higgs field, being the component of a gauge field,
is protected by radiative quadratic divergencies by the underlying higher-dimensional
gauge symmetry.

This idea has been first advocated in refs.\cite{early} but no concrete realization
was found. In particular, since the Higgs is a gauge field, the Yukawa couplings
are gauge couplings and thus it is not straightforward to get a realistic fermion
spectrum.
Interestingly enough, it has recently been understood that
realistic Yukawa couplings can be obtained in these models and in a manner
which provide a natural explanation of the large hierarchy of fermion
masses \cite{burdnom,CGM,Scrucca:2003ra}. This has allowed to construct several
interesting models of gauge-Higgs unification, both in a supersymmetric \cite{burdnom,Ghu-susy}
and in a non-supersymmetric context \cite{CGM,Scrucca:2003ra,Agashe:2004rs}.

The minimal model that one can consider is a five-dimensional theory
compactified on a segment (or $S^1/\Z_2$ orbifold) of length $L$, with gauge
group $G=SU(3)_c \times SU(3)_w$ \cite{Scrucca:2003ra}\footnote{In order to get
the correct weak-mixing angle, a further 5D $U(1)^\prime$ gauge field has to be introduced, but we can
neglect it in the considerations that will follow.}. If one suitably breaks
the $SU(3)_w$ gauge group down to $SU(2)_L\times U(1)_Y$ by appropriate orbifold
boundary conditions, one ends up with a 5D spectrum of Kaluza-Klein states,
in which the only massless fields (zero modes) are the 4D gauge bosons
$A_\mu^a$ ($a=1,2,3$) and $A_\mu$ of $SU(2)_L\times U(1)_Y$ ($\mu=0,1,2,3$)
and a complex scalar doublet $H$ coming from $A_5$, the Higgs field.
Gauge invariance forbids any local potential for $A_5$ in the interior of the segment
(bulk), the only allowed
gauge-invariant local operators being built with the field strength $F_{MN}$.
Actually, a remnant of the 5D $SU(3)_w$ gauge symmetry also forbids
any local potential for $A_5$ at the boundaries as well. In fact, at the
boundaries there is a symmetry acting non-linearly on the Higgs field \cite{vonGersdorff2}:
\be
\delta A_5 = \partial_5 \xi\,,
\label{nonlinear}
\ee
where $\xi$ are the gauge parameters
of $SU(3)_w/[SU(2)_L\times U(1)_Y]$.
The only gauge invariant operator that can give rise to a
Higgs potential $V(H)$ must then be non-local in the extra dimension
and expressed in term of the Wilson line $W={\cal P} \exp(i\int dy A_5)\equiv \exp(i\alpha)$, where
$0\leq \alpha \leq 2 \pi$ is the Wilson line phase \cite{hos}, related to the Higgs vacuum
expectation value $v$ by $\alpha\simeq v L$ (notice that $\alpha$ defined here differs
by a $2\pi$ factor from the $\alpha$ defined in ref.\cite{Scrucca:2003ra}).
The crucial and most important property of this construction
is that $V(H)$, being a function of $W$, is necessarily radiatively generated
and non-local in the extra dimension.
Being a non-local operator, $V(H)$ is finite at all orders in perturbation theory \cite{Non-local}.
No dependence on the UV cut-off $\Lambda$ appears in $V(H)$ and thus
the little hierarchy problem is solved. Depending on the field content
of the model, one could then have a radiatively induced electroweak symmetry breaking (EWSB),
governed by the Wilson line phase $\alpha$.\footnote{See respectively refs.\cite{WEffPot} and ref.\cite{Oda:2004rm} 
for studies of the structure of one-loop Wilson line potentials on flat and warped orbifolds.}
The EWSB is thus equivalent to a Wilson line symmetry breaking.

As we mentioned, the introduction of matter in this framework is not
straightforward. If one assumes that the SM fermions are fields localized
at the boundaries, then the symmetry (\ref{nonlinear}) forbids local couplings between
them and the Higgs field.
On the other hand, if they are 5D fields propagating
along the whole segment (bulk fields), their Yukawa couplings will necessarily
be all the same and given by the gauge coupling constant.
An interesting possibility to overcome this difficulty is obtained by
assuming that the SM fermions are localized fields with 
mixing terms with bulk massive fermions \cite{CGM,Scrucca:2003ra}. Since the bulk fermions couple to
the Higgs, thanks to the mixing, an effective Yukawa coupling will be induced
among the SM fermions. In fact, the effective Yukawa couplings between the
Higgs and the SM fermions achieved in this way are roughly given by \cite{Scrucca:2003ra}
\be
Y_f \sim \epsilon_L \, \epsilon_R \,
 g_4 \, LM_f\, e^{- L M_f}\,.
\label{yukawa}
\ee
In eq.(\ref{yukawa}), $M_f$ is the mass of the bulk fermion coupled with the
SM fermion $f$, $g_4$ is the 4D gauge coupling constant and $\epsilon_{L,R}$ are dimensionless couplings
which govern the mixing between the bulk fermion and the left- and right-handed SM
fermion $f$. Notice that the couplings $\epsilon_{L,R}$ are bounded, their values ranging
from 0 (no mixing with bulk fermions) to 1 (maximal mixing with bulk fermions).
The Yukawa couplings are effective (rather than fundamental)
couplings, which depend exponentially on $M_fL$. In this way, we may not only get
a realistic pattern of Yukawa couplings, but also have an understanding of
their hierarchy in terms of the exponential behaviour appearing in
eq.(\ref{yukawa}).\footnote{An alternative but essentially equivalent way of getting
exponentially suppressed Yukawa couplings is obtained by considering massive fermions on the segment.
In this case, a relation similar to eq.(\ref{yukawa}) is found, with $\epsilon_{L,R}=1$
(maximal mixing) \cite{burdnom}.}

Given this field content, one can thus compute the one-loop Higgs effective potential.
As we have already argued, this is necessarily finite.
One finds that an EWSB occurs with a value of the Wilson line phase $\alpha$ at the minimum
that is about $\sim 1/2\div 1$.
All the qualitative features of the SM are then nicely reproduced.
At the quantitative level, however, there are some problems.
Their occurrence can actually be predicted on general grounds and are somehow model independent:
\begin{itemize}
\item{$V(H)$ is radiatively generated. From a 4D perspective, one would expect a small Higgs quartic
coupling in general, leading to a too light Higgs mass.}
\item{The effective Yukawa couplings (\ref{yukawa}) are exponentially suppressed by $M_fL$.
This is fine for all the SM fermions, but the top quark. Unless some other mechanism is advocated,
one would expect from eq.(\ref{yukawa}) a too light top mass.}
\item{The compactification scale is determined by the value of the Wilson line phase $\alpha$ at
the minimum, since $M_W=\alpha/(2L)$. For $\alpha \sim 1/2 \div 1$, this results in a too low
compactification scale, given the current bounds (see \it{e.g.} \cite{Delgado:1999sv}).}
\end{itemize}
These problems can be solved, or alleviated, in various ways.
One possibility is to increase the value of the 5D gauge coupling constant $g_5$, which is the
microscopic coupling that governs the size of the Yukawa couplings and of the Higgs effective potential.
In flat space, $g_5$ is simply related
to the 4D coupling constant $g_4$ by the simple relation $g_5=g_4 \sqrt{L}$. Since $L$ and
$g_4$ are fixed by the experimental values of $M_W$ and of the $SU(2)_L$ SM gauge coupling constant,
the only way to increase $g_5$ is to introduce modifications in the model
that change the above relation between $g_4$ and $g_5$.
A simple way to do that is provided by adding kinetic terms for the 4D gauge fields $A_\mu$,
localized at the boundaries.
If these terms are large enough, their net effect is
to increase the Higgs and the top mass to realistic values \cite{Scrucca:2003ra}.
Since the relation between $M_W$ and $\alpha$ is also modified in presence
of localized gauge kinetic terms, it turns out that one can get
phenomenological acceptable values for the compactification scale as well.
All the above problems are solved, but unfortunately other potential problems
are introduced. They are all related to the fact that these localized
gauge kinetic terms introduce mixing among all Kaluza-Klein states.
This results in unwanted effects, such
as too large deviations to the $\rho$ parameter or to a non-universality of the 4D gauge
couplings \cite{Scrucca:2003ra}.
An other interesting way (probably closely related to the former) to increase the
5D coupling constant is obtained by considering a warped, rather than flat, space \cite{Contino:2003ve}.
In this case, the Higgs mass is generally higher than the value obtained in flat
space compactifications \cite{Agashe:2004rs,Hosotani:2005nz}, as well as the
Yukawa couplings, which are dynamically generated in a way that is essentially
the same as in flat space.
The warping, however, produces distortions similar to those
given by adding localized gauge kinetic terms in flat space.
By suitably imposing a custodial $SU(2)$ symmetry to the Higgs sector,
an interesting model of gauge-Higgs unification in warped space has been constructed
where the distortions might be under control and small enough to
be compatible with the Electroweak Precision Tests (EWPT) \cite{Agashe:2004rs}.\footnote{Interestingly
enough, the model of ref.\cite{Agashe:2004rs} has a purely 4D dual interpretation
as a composite Higgs model.}

An other possibility to solve the listed three problems is to find some microscopic
mechanism to dynamically stabilize the Wilson line phase $\alpha$ to a smaller value, such as
$5\times 10^{-2}$ or smaller. In this case, the Higgs quartic coupling is effectively enhanced
and can give rise to realistic Higgs masses. The compactification scale would also be
above the current bounds. The top mass problem is not directly solved in this way,
unless this new mechanism also allows for greater Yukawa couplings.
Unfortunately, there is no known satisfactory mechanism which allows to get values
of $\alpha\sim 5\times 10^{-2}$. It is interesting to note, however, that massive 5D fermions
in very large representations of the gauge group typically tend to give lower values
of $\alpha$ and also allows for bigger Yukawa couplings \cite{Scrucca:2003ra,Martinelli:2005ix}.
The representations needed are however very large, and would lead to a breakdown of an
effective field theory approach, since they lead to a NDA estimate of the cut-off
$\Lambda \sim 1/L$.

So far, we focused on one compact extra dimension, but what happens if one
has more extra dimensions ? Since the NDA estimate of $\Lambda$ decreases with
the number of extra dimensions and no new interesting features seem to appear
in further increasing their number, let us only consider the case
of two extra dimensions, namely a 6D theory.
In 6D, there are several potentially interesting two-dimensional compact spaces
one could consider. The simplest spaces, leading to a 4D chiral spectrum of fermions,
are given by orbifolds of tori of the form $T^2/\Z_N$, where $N=2,3,4,6$.
Let us focus on these spaces in the following.

There are two main important features that happen when going to 6D.
The first, good feature, is the appearing of a gauge-invariant Higgs quartic coupling at tree-level,
simply arising from the non-abelian part of the internal components of the
gauge field kinetic term $F_{56}^2$.
A tree-level quartic coupling is welcome, because it can automatically
solve the problem of a too light Higgs.
The second, bad feature, is the possible appearance of a local, gauge-invariant, operator
that contributes to the Higgs mass. This is an operator localized at the fixed-points of the
$T^2/\Z_N$ orbifold, with a quadratically divergent coefficient,
in general \cite{CGM,vonGersdorff2,SSSW,Biggio:2004kr}.
It is linear in the internal components of the field-strength $F$. Its abelian term
corresponds to a tadpole for certain gauge field components, whereas its non-abelian part
represents a mass term for the Higgs field. If there is no symmetry to get rid
of this operator, the hierarchy problem is reintroduced.
It turns out that a discrete symmetry forbidding this operator can be implemented
only for $T^2/\Z_2$ orbifolds, in which case, however, one gets two Higgs doublets, rather than one.
In this case, the Higgs effective potential has various similarities with the one
arising in the Minimal Supersymmetric Standard Model (MSSM). Explicit computations on
a given 6D model \cite{Antoniadis:2001cv} have shown that the lightest Higgs field turns
out to be again too light \cite{Hosotani:2004wv}.

Maybe a more interesting possibility is obtained by considering $T^2/\Z_N$ orbifolds, with $N\neq 2$.
If $N\neq 2$, one can get 2, 1 or 0 Higgs doublets, depending on the orbifold projection.
The most interesting case appears to be given by the 1 Higgs doublet models,
for which one finds $M_H = 2 M_W$ at tree-level, by geometrical considerations \cite{SSSW}.
However, no symmetry forbids the appearance of the localized operator mentioned above, which would
spoil the stabilization of the electroweak scale. Even if this operator is put to zero at tree-level,
no accidental one-loop cancellation seems to be possible. The best one can do is to advocate
a spectrum of 6D fields such that the sum of the one-loop quadratically divergent coefficients over all
fixed points vanish (global cancellation). In this case, it actually turns out that the
electroweak scale is not destabilized. Contrary to the 5D construction considered before,
the quadratic sensitivity to the cut-off would presumably be reintroduced at two-loop level, 
but a one-loop cancellation might
be enough to solve the little hierarchy problem. No concrete model has been yet presented
along these lines and thus it is premature to establish whether gauge-Higgs unification
in 6D can be a realistic proposal or not.


The idea of a Higgs field as a gauge boson in extra dimensions
seems to be a promising candidate to more conventional scenarios
of new physics, such as SUSY.

Several aspects of this idea require further study.
From a more theoretical side, it is desirable to find some mechanism
to increase the Higgs mass without introducing the unwanted distortion effects
that appears when one considers warped models or theories in flat space with
localized gauge kinetic terms.
It has also to be understood whether such theories (as many other theories
in extra dimensions) admit a microscopic completion where the orbifold singularities (for 6D models) or the
boundaries of the segment (for 5D models) are replaced by
an UV model defined on a smooth compact space \cite{sw}.

From a more phenomenological side, there are several issues which deserves
further study: the generic suppression of Flavour Changing Neutral Currents or
a systematic classification of all possible CP violating terms would be desirable.
The latter study would also shed light on the possibility of having Baryogenesis
at the electroweak scale, considering that a moderately strong first-order phase
transition can be obtained in these models \cite{Pan}.
It would also be interesting to better understand whether a possible Dark Matter candidate
can be found in such theories and under what conditions gauge coupling unification
(typically lost in these models) can be recovered (see ref.\cite{Agashe:2005vg} for a recent proposal).


I thank the hospitality of the Aspen Center for  Physics, where this work
has been completed.

\end{document}